\documentclass[11pt]{article}

\usepackage{amsmath}
\usepackage{graphicx}
\usepackage{indentfirst}
\usepackage{amssymb}
\usepackage{cite}
\usepackage{color}
\usepackage{subfigure}
\usepackage{kotex}

\begin{document}

\title{\bf{Geometrically Induced Semiclassical Limit for Quasinormal Modes and Greybody Factors in Higher-Dimensional Near-Ultracold spacetimes}}

\date{}
\maketitle

\begin{center}
\author{Bogeun Gwak}$^a$\footnote{rasenis@dgu.ac.kr}\\

\vskip 0.25in
$^{a}$\it{Department of Physics, Dongguk University, Seoul 04620, Republic of Korea}\\
\end{center}
\vskip 0.6in

{\abstract
{In this work, we investigate the correspondence between quasinormal modes (QNMs) and greybody factors (GBFs) in higher-dimensional Reissner--Nordst\"{o}m--de Sitter black holes based on the massless scalar field. Our analysis focuses on the QNMs in the near-ultracold family where the inner, outer, and cosmological horizons are close together. Under the geometric limit condition, the leading-order Wentzel--Kramers--Brillouin term of the QNM is dominant for fixed angular quantum numbers and dimensionality without imposing the eikonal limit. Particularly, the corresponding GBFs also retain the explicit form of both parameters. Furthermore, we demonstrate that the half-transmission frequency of the GBF and its derivative are related to null geodesics, in terms of aspects such as the angular frequency and Lyapunov exponent, with the use of both parameters. We provide their relationships in arbitrary dimensionality. Furthermore, the higher-order corrections to the QNMs are investigated at the Nariai limit and in the near-ultracold geometry with separated horizons.}}

\thispagestyle{empty}
\newpage
\setcounter{page}{1}

\section{Introduction}\label{sec1}

Quasinormal modes (QNMs) are damped oscillations in which the energy of the modes is dissipated to the asymptotic regime from a black hole\cite{Vishveshwara:1970zz,Press:1971wr}. According to the classical boundary, only ingoing QNMs exist at the event horizon. For the outgoing condition, the QNMs are only outgoing for asymptotically flat spacetime fields\cite{Kokkotas:1999bd}, reflected in anti-de Sitter (AdS) spacetimes\cite{Horowitz:1999jd}, and outgoing into the cosmological horizon in de Sitter (dS) spacetimes\cite{Cardoso:2003sw,Zhidenko:2003wq}. Even if the boundary conditions depend on the asymptotic geometries, the QNM plays a prominent role in the testing to structures of spacetimes because its waveform is encoded based on the effective potential, geometry, and boundary conditions resulting from interactions with a black hole or objects\cite{Cardoso:2016rao}. Moreover, black holes and wormholes possess different structures. Therefore, their effective potential and boundary conditions produce distinguishable QNM spectra\cite{Konoplya:2016hmd}. The boundary on outgoing waves is also closely related to the stability issue on a black hole referred to as superradiance\cite{Brito:2015oca}. In reflective boundaries, such as those of mirror and AdS spacetimes, the wave can be amplified by repeating the reflection when the outgoing wave cannot dissipate into the asymptotic regime. The amplified wave can make a black hole unstable like a bomb\cite{Cardoso:2004hs}.

Recently, a correspondence was established between QNMs and the greybody factor (GBF)\cite{Konoplya:2024lir}. The GBF is provided by different boundary conditions than those of QNMs. Therefore, they are physically distinct phenomena. These phenomena are technically expressed by a Schr\"{o}dinger-type equation with a single-maximum potential in which the Wentzel--Kramers--Brillouin (WKB) method\cite{Iyer:1986np} can demonstrate accurately wave behaviors. As independent QNM and GBF problems use identical effective potentials and the WKB method\cite{Konoplya:2003ii,Konoplya:2019hlu}, their relationship can be found as an approximate correspondence. The correspondence is also extended to axially symmetric black holes, and superradiance breaks the correspondence\cite{Konoplya:2024vuj}. Furthermore, QNMs and GBFs are connected by these types of correspondences in dS black holes and wormholes\cite{Malik:2024cgb,Bolokhov:2024otn,Lutfuoglu:2025hjy}. However, applying the WKB method on the GBF sometimes causes a divergence, which makes the correspondence invalid\cite{Pedrotti:2025idg}. When the dimensionality is lifted into higher-dimensional spacetime, there is an effective double-peak potential in which the correspondence becomes mostly invalid\cite{Han:2025cal,Han:2026fpn}. However, the breakdown of the correspondence depends on dimensionality and conserved quantities for electrically charged black holes\cite{Han:2026txd}.

The methods used to find a QNM spectrum are closely related to particular modes and the effective potential; correspondingly, such methods are mostly dependent on black holes. Our study focuses on an analytic solution to a QNM. From this perspective, the P\"{o}shl--Teller (PT) potential is also a good starting point because it provides an analytic solution for dS black holes in conjunction with the Nariai limit where the outer and cosmological horizons match\cite{Cardoso:2003sw,Molina:2003ff}. The PT potential also appears in accelerating black holes, such as those described by the C-metric\cite{Gwak:2022nsi}. In the case of the Schwarzschild-dS black hole, the first order of the PT potential does not contribute to the QNM frequency because of its oddness\cite{MaassenvandenBrink:2003yq}. Moreover, In the case of the Reissner--Nordstr\"{o}m--dS (RN--dS) black hole, analytic QNMs in the near-Nariai geometry case are identified, commonly referred to as photon-sphere QNMs\cite{Dias:2018etb}. By using numerical and analytic methods, QNMs in the cases of Kerr--Newman--dS (KN--dS) black holes are described in terms of analytic approximations\cite{Davey:2024xvd}. In Kerr--dS (K--dS) black holes, the ultracold limit where the three horizons are unified into a single horizon allows the derivation of specific QNM quantization conditions\cite{Arnaudo:2025btb}. 

In this study, we investigate the correspondence between QNMs and GBFs in the near-ultracold family in higher-dimensional RN--dS black holes based on the massless scalar field. As the leading-order QNM determined by the WKB method is dominant in the near-ultracold geometry, we determine the form of the QNMs without using the eikonal limit because the QNM becomes effectively a semiclassical limit in near-ultracold geometry. We apply this geometrical feature to the corresponding GBFs, which are also dominant in the leading order. This provides better resolution to analyze the correspondence in finite angular numbers because the first-order WKB is dominant in both cases when this geometry is used. Moreover, we assessed the QNMs and GBFs according to null geodesics. In the near-ultracold geometry, the relationships between QNMs (GBFs) and null geodesics are also enhanced to form a finite angular number instead of the eikonal limit. Based on our analysis, the near-ultracold geometry provides the geometrically semiclassical limit instead of the eikonal limit. Thus, the relationships from the leading-order WKB method become dominant in this geometry case. To explore the near-ultracold family, we determine the Nariai limit in the family. By showing that the QNMs change into that of the PT potential, we validate our results and comment on the higher order of the QNMs. Furthermore, we determine whether such geometric effects continue to act on QNMs far from near-ultracold geometry. The effect is found to be unified only in the near-ultracold geometry case and separated into geometric and angular effects far from the ultracold geometry. We determine all of these relationships in terms of arbitrary dimensions.

The remainder of this paper is organized as follows. We briefly review higher-dimensional RN--dS black holes in Section 2. The massless scalar field equation is then derived in arbitrary dimensions in Section 3. In Section 4, we determine QNMs and their GBF correspondences in the near-ultracold geometry case. The QNMs and GBFs are related to null geodesics in Section 5. In Section 6, we show that QNMs extend to the Nariai limit and are far from ultracold regimes; we also present a discussion on higher-order effects. We summarize our results in Section 7.

\section{Higher-dimensional RN--dS Black Holes}

We herein consider higher-dimensional RN--dS black holes, including that of four dimensions. They form a $D$-dimensional spacetime with a positive cosmological constant $\Lambda$. The metric is given by
\begin{eqnarray}\label{eq:metric01}
ds^2=-f(r)dt^2+f(r)^{-1}dr^2+r^2d\Omega_{D-2},\quad f(r)=1-\frac{2M}{r^{D-3}}+\frac{Q^2}{r^{2D-6}}-\frac{2\Lambda r^2}{(D-1)(D-2)},
\end{eqnarray}
where the analytic form of the $(D-2)$-dimensional sphere is described as\cite{Gwak:2017kkt,Gwak:2019asi}
\begin{align}\label{eq:metric02}
d\Omega_{D-2}^2=\sum_{i=1}^{D-2}\left(\prod_{j=1}^{i}\sin^2\theta_{j-1}\right)d\theta_i^2,\quad \theta_0=\frac{\pi}{2},\quad \theta_{D-2}=\phi.
\end{align}
Note that the azimuthal angle is $\phi$. The mass and electric charge parameters $M$ and $Q$ are related to the mass and electric charge of the black hole\cite{Tian:2010gn,Gwak:2019asi}. 
\begin{align}
M_\text{b}=\frac{(D-2)\Omega_{D-2}}{8\pi}M,\quad Q_\text{b}=\frac{(D-2)\Omega_{D-2}}{8\pi}Q.
\end{align}
The RN--dS black hole in Eq.\,(\ref{eq:metric01}) has three horizons referred to as the inner, outer, and cosmological horizons.
\begin{eqnarray}
r_\text{i}\leq r_\text{o}\leq r_\text{c},\quad f(r_\text{i})=f(r_\text{o})=f(r_\text{c})=0,
\end{eqnarray}
of which the locations can be determined by the metric function $f(r)$ in Eq.\,(\ref{eq:metric01}). As there are three horizons, their combinations also provide particular geometries with specific names. When $r_\text{i}=r_\text{o}$, the spacetime becomes an extremal black hole. The Nariai geometry is represented when $r_\text{o}=r_\text{c}$\cite{Gwak:2019ttv,Ko:2023nim}. Specifically, there is a geometry called ultracold spacetime when all three horizons are coincident\cite{Romans:1991nq,Dias:2004px}. In that case, this geometry has a triple root denoted as $r_\text{t}$, referred to as the ultracold horizon.
\begin{eqnarray}
r_\text{i}=r_\text{o}=r_\text{c}=r_\text{t},
\end{eqnarray}
where the ultracold geometry satisfies
\begin{align}\label{eq:ultracoldequations15}
f(r_\text{t})=0,\quad f'(r_\text{t})=0,\quad f''(r_\text{t})=0.
\end{align}
By using a combination of Eq.\,(\ref{eq:ultracoldequations15}), the mass and charge parameters and the cosmological constant can be rewritten in terms of $r_\text{t}$.
\begin{align}\label{eq:rmrt02}
M_\text{t}=\frac{2r_\text{t}^{D-3}}{D-1},\quad Q_\text{t}^2=\frac{r_\text{t}^{2D-6}}{D-2},\quad \Lambda_\text{t}=\frac{(D-3)^2}{2r_\text{t}^2}.
\end{align}
This shows that the ultracold geometry is unique for specific $D$ and $r_\text{t}$ values. Furthermore, for a fixed value of $\Lambda$, there is only one point corresponding to the ultracold geometry in the phase space $(M,Q)$, and the near-ultracold family forms a regime around this point.

We mainly considered the near-ultracold geometry in which the three horizons approach the ultracold horizon. The root-factorized form of the metric function $f(r)$ is technically convenient to treat the geometry.
\begin{align}\label{eq:metric05}
f(r)=\frac{(r-r_{\text i})(r-r_{\text o})(r-r_{\text c})G(r)}{r^{2D-6}},\quad G(r)=-\frac{2\Lambda}{(D-1)(D-2)}\prod_{k=1}^{2D-7}\left(r-r_k\right),
\end{align}
where $r_k$ are negative real or complex roots.

\section{Scalar Field Equation in Higher-dimensional RN--dS Black Holes}

We consider herein the massless scalar field in the higher-dimensional RN--dS black hole cases to determine a general formula by using a simple model. The scalar field acts as a probe field in a fixed background geometry, and we impose QNM boundary conditions. Hence, it decays according to time evolution. The action of the massless scalar field $\Psi$ is
\begin{align}
S_\Psi = -\frac{1}{2}\int d^{D} \sqrt{-g} g^{\mu\nu}\partial_\mu \Psi \partial_\nu \Psi.
\end{align}
The scalar field equation is
\begin{align}\label{eq:scalareq21}
\frac{1}{\sqrt{-g}}\partial \left(\sqrt{-g} g^{\mu\nu} \partial \Psi\right)=0,
\end{align}
where the metric determinant from Eqs.\,(\ref{eq:metric01}) and (\ref{eq:metric02}) is obtained as
\begin{align}
\sqrt{-g}=r^{D-2}\prod_{j=0}^{D-3}\sin^{D-2-j}\theta_j.
\end{align}
The metric in Eq.\,(\ref{eq:metric01}) has translation symmetries for the $t$ and $theta_{D-2}=phi$ coordinates. Hence, the ansatz for the scalar field equation can be constructed as
\begin{align}\label{eq:scalareq22}
\Psi(t,r,\Omega_{D-2})=A e^{-i\omega t} R(r) Y_{\ell m} (\Omega_{D-2}),
\end{align}
where $Y_{\ell m}$ is the hyperspherical harmonics defined in higher-dimensional spacetime with a $(D-2)$-spherical symmetry, and the $\phi$-directional ansatz is absorbed into $Y_{\ell m}$. Note that $\ell$ and $m$ are the angular quantum numbers for each $\Omega_{D-2}$ direction. Substitution of Eq.\,(\ref{eq:scalareq22}) in Eq.\,(\ref{eq:scalareq21}) becomes separable into equations along each coordinate direction. The $\Omega_{D-2}$-directional equations are written in a recurrent form with the use of hyperspherical harmonics\cite{Ida:2002zk}.
\begin{align}
\nabla_{S^{D-1-i}}^{2} Y_{lm}=(\sin\theta_i)^{2+i-D}\frac{\partial}{\partial\theta_i}\left[(\sin\theta_i)^{D-2-i}\frac{\partial}{\partial\theta_i}Y_{lm}
\right]+\frac{1}{\sin^2\theta_i}\nabla_{{S^{D-2-i}}}^{2}Y_{lm} \,\, (i=1,2,\ldots,D-3),
\end{align}
where $\nabla_{{S^{i}}}^{2} $ is the Laplace--Beltrami operator with respect to $S^{i}$. Thus, the separation constant becomes the total angular quantum number of the scalar field.
\begin{align}
\nabla^2_{S^{D-2}} Y_{\ell m}=-\ell(\ell+D-3)Y_{\ell m}.
\end{align}
By the separation constant, the radial equation becomes\cite{Konoplya:2003ii}
\begin{align}\label{eq:radialeq01}
\frac{1}{r^{D-2}}\partial_r \left(r^{D-2}f(r)\partial_rR(r)\right)+\left(\frac{\omega^2}{f(r)}-\frac{\ell(\ell-D-3)}{r^2}\right)R(r)=0.
\end{align}
Based on our analysis, the QNM frequencies can be determined from the radial equation. Therefore, we focus on solving the radial equation. To rewrite the radial equation to a Schr\"{o}dinger-type equation, we redefine and impose tortoise coordinates.
\begin{align}
R(r)=r^{-\frac{D-2}{2}}\psi(r),\quad \frac{dr_*}{dr}=\frac{1}{f(r)}. 
\end{align}
Accordingly, Eq.\,(\ref{eq:radialeq01}) becomes a Schr\"{o}dinger-type equation.
\begin{align}\label{eq:radialeq05}
\frac{d^2 \psi}{dr_*^2}+\left(\omega^2-V(r)\right)\psi=0,
\end{align} 
where the effective potential is
\begin{align}\label{eq:radialeq07}
V(r)=f(r)\left(\frac{\ell(\ell+D-3)}{r^2}+\frac{(D-2)f'(r)}{2r}+\frac{(D-2)(D-4
)f(r)}{4r^2}\right),
\end{align}
and
\begin{align}
f'(r)=-\frac{(2D-6)Q^2}{r^{2D-5}}+\frac{2(D-3)M}{ r^{D-2}}-\frac{4r\Lambda}{(D-2)(D-1)}.
\end{align}
Beginning with Eqs.\,(\ref{eq:radialeq05}) and (\ref{eq:radialeq07}), we consider the near-ultracold family of RN--dS black holes in higher dimensions. Note that $\ell$ and $D$ are considered finite in our analysis.

\section{Correspondence of QNMs and GBFs}
\label{sec:qnm_gbf_correspondence}

We determine the near-ultracold limit of QNMs in higher dimensions. GBFs are then produced from the QNMs using the correspondences between them\cite{Konoplya:2024lir}. They are actually determined by different boundary conditions. QNMs are not incident waves at both horizons. GBFs are the transmission probabilities when the incident wave has a real frequency. However, they share the same effective potential data in the leading-order WKB method. Based on this relationship, GBFs can be constructed from QNM data. This is called the correspondence between QNMs and GBFs.

\subsection{QNMs in Geometric Limit}

We investigate QNMs of the scalar field in the higher-dimensional RN--dS black hole. The QNM boundary conditions are imposed as
\begin{align}
\psi(r_*)&\sim e^{-i\omega r_*},\quad r_*\rightarrow-\infty,\\
\psi(r_*)&\sim e^{+i\omega r_*},\quad r_*\rightarrow+\infty.\nonumber
\end{align}
The outer and cosmological horizons behave like classical boundaries; therefore, the ingoing and outgoing waves are dissipated into horizons according to the set conditions. We systematically treat the near-ultracold black hole by defining geometric variables.
\begin{align}\label{eq:nearultracoordinate21}
r_\text{m}=\frac{r_\text{o}+r_\text{c}}{2},\quad\epsilon=\frac{r_\text{m}-r_\text{i}}{r_\text{t}},\quad\sigma=\frac{r_\text{c}-r_\text{o}}{2\left(r_\text{m}-r_\text{i}\right)},\quad\rho=\frac{r-r_\text{m}}{\sigma\epsilon r_\text{t}},
\end{align}
where $r_\text{m}$ is the middle point between the outer and cosmological horizons. $\epsilon$ is the configuration parameter determining how far x is from the ultracold limit. $\sigma$ is the horizon-gap ratio showing whether the geometry closes the ultracold or Nariai geometry. $\rho$ is the dimensionless radial coordinate instead of the $r$ coordinate. 

We consider herein a particular near-ultracold family, with $\epsilon\ll 1$ and finite $\sigma$. Hence, all the horizons are close to $r_\text{m}$; however, the relative ratios between horizons are maintained finite. In combinations of Eq.\,(\ref{eq:rmrt02}) and Eq.\,(\ref{eq:metric05}), three horizons provide
\begin{align}
r_{\text m}=r_{\text t}+\frac{1}{3}r_{\text t}\epsilon-\frac{1}{36}(3D-11)(1+3\sigma^2)r_{\text t}\epsilon^2+\mathcal{O}(\epsilon^3).
\end{align}
Accordingly, we obtained the second-order term for convenience and thus considered the expansion up to the first-order term. Note that there is only one ultracold state for a given cosmological constant. Thus, $r_\text{t}$ has a unique value for a given $\Lambda$ family. Hence, when $\epsilon$ is finite, $(M,Q)$ forms an area in the parameter space. Furthermore, $f(r)$ can be expanded to
\begin{align}
f(\rho)=\frac{2(D-3)^2}{3}\sigma^2\epsilon^3P_\sigma(\rho)+\cdots,\quad P_\sigma(\rho)\equiv(1-\rho^2)(1+\sigma\rho).
\end{align}
To obtain dimensionless variables, we impose transformations such as
\begin{align}\label{eq:rescalexx01}
\bar{\kappa}\equiv\frac{\kappa_o+\kappa_c}{2},\quad
\tilde{\omega}\equiv\frac{\omega}{\bar{\kappa}},\quad x\equiv\bar{\kappa}r_*,\quad  W\equiv\frac{V}{\bar{\kappa^2}},\quad \tilde{V}\equiv k_\text{eff}^2\frac{V}{\bar{\kappa^2}},
\end{align}
where $\bar{\kappa}$ is the average of the surface gravities on the outer and cosmological horizons. Based on Eq.\,(\ref{eq:rescalexx01}), the tortoise coordinate is also given as
\begin{align}
\frac{dx}{d\rho}=\frac{\bar{\kappa}\epsilon\sigma r_\text{t}}{f(\rho)}.
\end{align}
Using Eq.\,(\ref{eq:radialeq05}), we can determine the Schr\"{o}dinger-type equation.
\begin{align}
\left(-h_{\text{eff}}^2\frac{d^2}{dx^2}+\tilde{V}(\rho)\right)\psi=h_{\text{eff}}^2\tilde{\omega}^{2}\psi,
\end{align}
where
\begin{align}\label{eq:v0potential01}
h_{\text{eff}}^2\equiv\frac{2(D-3)^2\epsilon}{3\ell(\ell+D-3)},\quad \tilde{V}\equiv P_\sigma(\rho)+\mathcal{O}(\epsilon).
\end{align}
This equation already shows what we mean by geometric limit. In the eikonal limit, $\ell \gg 1$, the value of $k_\text{eff}$ becomes small, and the WKB method becomes exact. However, without the eikonal limit, we can find exact QNM frequencies with a finite value of $\ell$ by taking a geometric limit of $\epsilon \ll 1$. Again, when the black hole becomes near-ultracold, that is, $\epsilon\ll 1$, the value of $k_\text{eff}$ becomes very small; thus, the WKB method produces an exact frequency with a finite value of $\ell$.

We applied the WKB method to obtain the QNM frequency under the geometric limit condition. From Eq.\,(\ref{eq:v0potential01}), the effective potential has a maximum, and its maximum location is obtained as
\begin{align}\label{eq:peakrhop31}
\rho_\text{p}=\frac{-1+\sqrt{1+3\sigma^2}}{3\sigma},\quad P_{\sigma} (\rho_\text{p})\equiv P_{\sigma\text{p}},
\end{align}
and
\begin{align}
\tilde{V}_\text{p}\equiv\tilde{V}|_{x=x_\text{p}}=P_{\sigma\text{p}},\quad \tilde{V}''_\text{p}\equiv \frac{d^2\tilde{V}}{dx^2}|_{x=x_\text{p}}=-2P_{\sigma\text{p}}^2\sqrt{1+3\sigma^2}.
\end{align}
By using the WKB method, we can determine the analytic form of the QNM frequency when $k_\text{eff}\ll 1$.
\begin{align}\label{eq:qnmunltracold01}
\tilde{\omega}_{n,\pm}=\pm\frac{\sqrt{P_{\sigma\text p}}}{h_{\text{eff}}}-i\sqrt{P_{\sigma\text p}}\left(1+3\sigma^2\right)^{1/4}\left(n+\frac{1}{2}\right)+\mathcal{O}\left(h_{\text{eff}}\right).
\end{align}
As the potential is real and frequency-independent, there are two branches for the QNM frequencies. Their relationship is found as
\begin{align}
\tilde{\omega}_{n,-}=-\tilde{\omega}_{n,+}^*.
\end{align}
For near-ultracold RN--dS black holes, we can find the exact QNM frequency with a finite value of $\ell$ because this geometric limit makes the WKB method exact. This can be achieved by
\begin{align}
\epsilon\rightarrow 0\quad\Longrightarrow\quad h_{\text{eff}}\rightarrow 0.
\end{align}
When the three horizons are geometrically coincident to one ultracold horizon, the value of $k_\text{eff}$ with a finite $\ell$ goes to zero without the eikonal limit. Therefore, the QNM frequency has the exact form for a given finite value of $\ell$ on near-ultracold RN--dS black holes.

\subsection{Leading-order Correspondence}

The real frequency of the incident wave in GBFs is given by
\begin{align}
\tilde{\Omega}&\equiv\frac{\Omega}{\bar{\kappa}},
\label{eq:gbf_frequency_normalization}
\end{align}
where $\tilde{\Omega}$ is the value normalized by the average surface gravity. As our QNM data in Eq.\,(\ref{eq:qnmunltracold01}) refer to the leading-order WKB method, the GBF, denoted by $\Gamma_\ell$, is also obtained from the leading-order QNM-GBF correspondence given in \cite{Konoplya:2024lir}.
\begin{align}
\Gamma_\ell(\Omega)=\left(1+\exp\left(\frac{\pi\left(\tilde{\Omega}^2
-(\text{Re}\tilde{\omega}_{0,+})^2\right)}{2\text{Re}\tilde{\omega}_{0,+}\text{Im}\tilde{\omega}_{0,+}}\right)
\right)^{-1}+\mathcal{O}(\epsilon^{1/2}),
\label{eq:gbf_correspondence_qnm}
\end{align}
where the QNM data are expressed using normalized variables. Note that the overtone number $n=0$ needs the correspondence in Eq.\,(\ref{eq:gbf_correspondence_qnm}). Then,
\begin{align}
\operatorname{Re}\tilde{\omega}_{0,+}=\frac{\sqrt{P_{\sigma\text{p}}}}{h_{\text{eff}}}+\mathcal{O}(\epsilon^{1/2}),\quad \operatorname{Im}\tilde{\omega}_{0,+}=-\frac{1}{2}\sqrt{P_{\sigma\text{p}}}(1+3\sigma^2)^{1/4}+\mathcal{O}(\epsilon^{1/2}).
\label{eq:gbf_fundamental_input}
\end{align}
Note that the values of $D$ and $\ell$ are still arbitrary, and $h_{\text{eff}}=\mathcal{O}(\epsilon^{1/2})$. Then, the correspondence produces
\begin{align}
\Gamma_\ell(\Omega)=\left(1+\exp\left(-\frac{\pi h_{\text{eff}}}{P_{\sigma \text{p}}(1+3\sigma^2)^{1/4}}\left(\tilde{\Omega}^{2}-\frac{P_{\sigma\text{p}}}{h_{\text{eff}}^2}\right)\right)\right)^{-1},
\label{eq:gbf_leading}
\end{align}
where $h_{\text{eff}}>0$, and $P_{\sigma \text{p}}>0$; therefore, $0<\Gamma_\ell<1$. The higher frequency of the incident wave is the higher transmission probability. The correspondence has a better accuracy in the eikonal limit $\ell\gg 1$. However, the GBF in Eq.\,(\ref{eq:gbf_leading}) is accurate when a finite $\ell$ value is used because the geometric limit $\epsilon \ll 1$ makes $k_\text{eff} \ll 1$. Thus, this is the finite-$\ell$ GBF in near-ultracold geometry.

\subsection{Half Transmission and Derivative on GBF}

The GBF in Eq.\,(\ref{eq:gbf_leading}) can be tested by using half transmission and its derivative, and these properties play a prominent role in the verification by geodesics. The half-transmission frequency can be simply obtained from Eq.\,(\ref{eq:gbf_leading}). Then, this matches the real part of the QNM with $n=0$ in Eq.\,(\ref{eq:qnmunltracold01}).
\begin{align}
\left.\tilde{\Omega}\right|_{\Gamma_\ell=1/2}=\frac{\sqrt{P_{\sigma\text{p}}}}{h_{\text{eff}}}=\operatorname{Re}\tilde{\omega}_{0,+}+\mathcal{O}(\epsilon^{1/2}).
\label{eq:gbf_half_frequency}
\end{align}
The half-transmission frequency has a close relationship with the real part of the QNM.

The derivative with respect to the incident frequency is related to the change in the GBF. When the incident wave is at the half-transmission frequency, the derivative on the GBF is 
\begin{align}
\left.\frac{d\Gamma_\ell}{d\tilde{\Omega}}\right|_{\Gamma_\ell=1/2}=\frac{\pi}{2\sqrt{P_{\sigma \text{p}}}(1+3\sigma^2)^{1/4}}+\mathcal{O}(\epsilon).
\label{eq:gbf_slope_normalized}
\end{align}
In comparison with Eq.\,(\ref{eq:qnmunltracold01}), the imaginary part of the QNM is connected into the derivative. Then,
\begin{align}
\left.\frac{d\Gamma_\ell}{d\tilde{\Omega}}\right|_{\Gamma_\ell=1/2}=
-\frac{\pi}{4\mathrm{Im}\,\tilde{\omega}_{0,+}}+\mathcal{O}(\epsilon^{1/2}).
\label{eq:gbf_derivative_normalized}
\end{align}
Hence, according to the leading-order correspondence, the GBF has a relationship with the real and imaginary parts on the QNM and with the half-transmission frequency and its derivative. Particularly, the half-transmission frequency is given by a finite-$\ell$ form in the near-ultracold limit, and the derivative is independent of $\ell$. We determined these relationships from the first-order WKB method; therefore, no further verification is required.

\section{Geodesic Validations on Correspondence}
\label{sec:geodesic_test}

We validate our results on QNMs and GBFs produced from the correspondence of the null geodesic angular frequency and Lyapunov exponent. This validation is independent of what we obtained from the WKB method; therefore, this process ensures the relationship between QNMs and GBFs.

\subsection{Null Geodesics}

The null geodesics are considered on the equator with $\theta=\frac{\pi}{2}$. From the metric in Eq.\,(\ref{eq:metric01}), the null condition and conserved quantities are obtained as
\begin{align}
-f(r)\dot{t}^{2}+\frac{\dot{r}^{2}}{f(r)}+r^2\dot{\phi}^{2}=0,\quad E=f(r)\dot{t},\quad J=r^2\dot{\phi},
\label{eq:geo_conserved_quantities}
\end{align}
where the energy and angular momentum of the null particle are $E$ and $J$, respectively. The effective potential is given by\cite{Cardoso:2008bp}
\begin{align}
V_r(r)=E^2-\frac{J^2f(r)}{r^2}.
\label{eq:geo_radial_potential}
\end{align}
The angular frequency and Lyapunov exponent are measured on an unstable circular orbit. In that case, the radius is constant; herein, it is denoted as $r_\text{ph}$. The conditions for the unstable circular orbit are
\begin{align}
V_r(r_{\text{ph}})=0, \quad V_r'(r_{\text{ph}})=0,\quad V_r''(r_{\text{ph}})>0. 
\end{align}
Note that the second-derivative term has a $(+)$ sign in our definition. The radius of the circular orbit is determined as
\begin{align}
r_{\text{ph}}=\left(\frac{1}{2}\left((D-1)M+\sqrt{(D-1)^2M^2-4(D-2)Q^2}\right)\right)^{\frac{1}{D-3}}.
\label{eq:geo_photon_radius_exact}
\end{align}
There is no cosmological constant; however, $r_\text{ph}$ still depends on $\Lambda$ because $M$ and $Q$ should be inside the allowed regime for a given $\Lambda$. Moreover, $r_o<r_{\text{ph}}<r_c$와 $f(r_{\text{ph}})>0$ should be satisfied to form the orbit. Based on the near-ultracold coordinate in Eq.\,(\ref{eq:nearultracoordinate21}), the circular orbit is written as
\begin{align}
\rho_{\text{ph}}&=\rho_\text{p}+\mathcal{O}(\epsilon),\quad r_{\text{ph}}=r_\text{t}+r_\text{t}\left(\frac{1}{3}+\sigma\rho_\text{p}\right)\epsilon+\mathcal{O}(\epsilon^2),
\label{eq:geo_photon_radius_expansion}
\end{align}
where $\rho_\text{p}$ is the same value as that we had determined for QNMs in Eq.\,(\ref{eq:peakrhop31}); however, it is an independent result from null geodesics. In leading order, $\rho_\text{ph}$ and $\rho_\text{p}$ are identical; however, they differ in finite-$\epsilon$ cases.

\subsection{QNMs and GBFs on Unstable Circular Trajectory}

The QNM frequency is complex, and its real and imaginary parts separately correspond to null geodesics in the higher dimensional RN--dS black hole case. The real part of the QNM frequency is related to the angular frequency of the null geodesics. Then,
\begin{align}
\tilde{\Omega}_{\text{ph}}=\frac{1}{\bar{\kappa}}\left.\frac{d\phi}{dt}\right|_{r=r_{\text{ph}}}=\frac{\sqrt{f(r_{\text{ph}})}}{\bar{\kappa}r_{\text{ph}}}=\frac{\sqrt{P_{\sigma\text{p}}}}{h_{\text{eff}}\sqrt{\ell(\ell+D-3)}}+\mathcal{O}(\epsilon^{1/2}),
\label{eq:geo_orbital_frequency_exact}
\end{align}
where the angular frequency is independent of $\ell$; however, we impose $k_\text{eff}$ with $\sqrt{\ell(\ell+D-3)}$. In comparison with the real part of the QNM frequency,
\begin{align}
\operatorname{Re}\tilde{\omega}_{n,+}&=\sqrt{\ell(\ell+D-3)}\tilde{\Omega}_{\text{ph}}
+\mathcal{O}(\epsilon^{1/2})=\frac{\sqrt{P_p}}{h_{\text{eff}}}+\mathcal{O}(\epsilon^{1/2}).
\label{eq:qnm_geo_real_test}
\end{align}
The real part of the QNM frequency is different from the angular frequency of the null circular orbit as much as the angular eigenvalue $\ell(\ell+D-3)$. This is similar to the relationship between QNMs and null geodesics in the eikonal limit. However, instead of $\ell \gg 1$, the semiclassical limit is achieved geometrically when $\epsilon \ll 1$. Thus, $\ell(\ell+D-3)$ connects both sides.

The imaginary part of the QNM is related to the Lyapunov exponent; therefore, it matches the decay rate. The Lyapunov exponent of null geodesics is given by the second derivative of the effective potential\cite{Cardoso:2008bp}.
\begin{align}
\lambda_{\text{ph}}=\left.\sqrt{\frac{V_r''(r)}{2\dot{t}^{2}}}\right|_{r=r_{\text{ph}}},
\label{eq:geo_lyapunov_definition}
\end{align}
where $V_r''(r_\text{ph})$ is positive by definition. The Lyapunov exponent is then obtained in the leading order as
\begin{align}
\tilde{\lambda}_{\text{ph}}\equiv\frac{\lambda_{\text{ph}}}{\bar{\kappa}}=\sqrt{P_{\sigma\text{p}}}(1+3\sigma^2)^{1/4}+\mathcal{O}(\epsilon).
\label{eq:geo_lyapunov_normalized}
\end{align}
The Lyapunov exponent is identical to the decay rate of the QNM with overtone number. Accordingly, we found that
\begin{align}
\text{Im}\,\tilde{\omega}_{n,+}
&=-\left(n+\frac{1}{2}\right)
\tilde{\lambda}_{\text{ph}}
+\mathcal{O}(\epsilon^{1/2}),
\label{eq:qnm_geo_imaginary_test}
\end{align}
where the $(-)$ sign of the Lyapunov exponent is owing to our sign choice of the effective potential. Based on these geodesic derivations of the QNM parts, we can validate our QNM results and maintain the finite $\ell$ value determined by the geometric limit.

The QNM frequency is shown to be closely related to the null geodesics, and the geometric limit in the near-ultracold regime is also important to maintain a finite $\ell$ value. Herein, we compare GBFs with the null geodesic properties. The near-ultracold limit still plays a major role as a geometric limit instead of the eikonal limit. Comparing Eq.\,(\ref{eq:gbf_half_frequency}) with Eq.\,(\ref{eq:geo_orbital_frequency_exact}), it is shown that the half transmission is related to the angular frequency with an angular quantum number\cite{Pedrotti:2025idg}.
\begin{align}
\left.\tilde{\Omega}\right|_{\Gamma_\ell=1/2}
&=\sqrt{\ell(\ell+D-3)}\,
\tilde{\Omega}_{\text{ph}}
+\mathcal{O}(\epsilon^{1/2}),
\label{eq:gbf_geo_center_test}
\end{align}
which corresponds to the leading term of the angular frequency with an angular quantum number. The reasoning for this finding is the same as that for the QNM case. Instead of the eikonal limit, the relation maintains the finite $\ell$ value in the near-ultracold limit.

Furthermore, we also found that the derivative of the GBF is related to the Lyapunov exponent. The derivative can be written as
\begin{align}\label{eq:gbf_slope_lyapunov_test}
\left.
\frac{d\Gamma_\ell}{d\tilde{\Omega}}\right|_{\Gamma_\ell=1/2}= \frac{\pi}{2\tilde{\lambda}_{\text{ph}}}+ O(\epsilon).
\end{align}
This relationship implies that the derivative is proportional to the inverse of the Lyapunov exponent. A smaller Lyapunov exponent yields a larger derivative at the half transmission of the GBF.

\section{QNMs Near-ultracold Family}

We previously defined the near-ultracold family based on the $(\epsilon,\sigma)$ variables. By using these variables, our QNM analysis can be extended to additional solutions given in the RN--dS black hole cases, such as those of the Nariai limit and finite-$\epsilon$ black hole. Under these geometries, we can evaluate the validity of the QNM solution in the near-ultracold limit and determine a finite-$\epsilon$ effect on the higher-order contribution.

\subsection{QNMs in Nariai Limit}

We can move the Nariai geometry in the near-ultracold family by setting $\sigma \ll 1$, according to which the effective potential becomes a well-known QNM frequency according to the PT potential. Furthermore, this can be only achieved by controlling $\sigma$; therefore, this connection can clarify the relationship between QNMs in the ultracold and Nariai limits and validate Eq.\,(\ref{eq:qnmunltracold01}). We can then choose the Nariai limit in the near-ultracold family by
\begin{align}
\epsilon\ll1,\,\,\text{and}\,\,\sigma\ll1.
\end{align}
This means that the three horizons are still close to each other; however, the outer and cosmological horizons are much closer than the inner one. This case also corresponds to the Nariai geometry. In the $\sigma\rightarrow 0$ limit, the tortoise coordinate becomes
\begin{align}
\frac{dx}{d\rho}&=\frac{1}{1-\rho^2},\quad \rho=\tanh x.
\end{align}
Based on the tortoise coordinate, the effective potential becomes the well-known PT-type potential that has a well-known solution form\cite{Cardoso:2003sw,Molina:2003ff,Gwak:2022nsi}.
\begin{align}
\tilde{V}_\text{PT}(x)=\text{sech}^2 x.
\end{align}
We rewrite the field equation as
\begin{align}
\frac{d^2\psi}{dx^2}+\left(\tilde{\omega}^{2}-\frac{h_{\text{eff}}^{-2}}{\text{cosh}^2 x} \right)\psi=0,
\end{align}
which refers to the PT potential; thus, the QNM frequency is
\begin{align}
\tilde{\omega}^{\text{PT}}_{n,\pm}=\pm\sqrt{\frac{1}{h_{\text{eff}}^2}-\frac{1}{4}}-i\left(n+\frac{1}{2}\right),
\end{align}
which is the exact PT potential form obtained without the use of the WKB method. We are still within geometric limits with $\sigma \rightarrow 0$ when $\epsilon\rightarrow 0$; thus, $h_{eff}\ll 1$. Accordingly, the leading order of $k_\text{eff}\ll 1$ in the QNM frequency is determined as 
\begin{align}\label{eq:qnmptcase01}
\tilde{\omega}^{\text{PT}}_{n,\pm}=\pm\frac{1}{h_{\text{eff}}}-i\left(n+\frac{1}{2}\right)+\mathcal{O}\left(h_{\text{eff}}\right).
\end{align}
When $\sigma\rightarrow 0$, the QNM frequency obtained using the WKB method in Eq.\,(\ref{eq:qnmunltracold01}) exactly reduces to that obtained using the PT-potential method in Eq.\,(\ref{eq:qnmptcase01}). Furthermore, there is no linear $\sigma$ term in the frequency because of the reflection symmetries on $x$ and $\sigma$. When $\sigma$ changes,
\begin{align}
P_{-\sigma}(-\rho)=P_\sigma(\rho).
\end{align}
Therefore, the tortoise coordinate also satisfies
\begin{align}
x_{-\sigma}(-\rho)=-x_\sigma(\rho),
\end{align}
where the QNM boundary conditions are also preserved. This combination implies/infers that the QNM frequency is even based on the assumption that the branch is simple and analytic near $\sigma=0$.
\begin{align}
\tilde{\omega}_n(\sigma)=\tilde{\omega}_n(-\sigma).
\end{align}
Thus, the odd $\sigma$-order frequency terms vanish.
\begin{align}
\tilde{\omega}_n^{(1)}=\tilde{\omega}_n^{(3)}=\tilde{\omega}_n^{(5)}=\cdots=0.
\end{align}

\subsection{Leading Geometric Semiclassical Effect in Finite-$\epsilon$ Regime}

We studied the QNM frequencies in the near-ultracold geometry case and clarified their continuous connection to those in Nariai geometry in the small $\epsilon$ family. Particularly, limiting zero of $k_\text{eff}$ geometrically defines the semiclassical regime. For this behavior, $k_\text{eff}$ should be proportional to $\epsilon^{1/2}/\ell$. Herein, we investigate whether $k_\text{eff}$ still defines the semiclassical regime (or not) for a finite value of $\epsilon$. This will determine the range of the geometrically semiclassical regime in terms of $\epsilon$.

Beginning with Eq.\,(\ref{eq:radialeq07}), we considered all relationships and counted up to the first order to clarify the effect of the finite value of $\epsilon$. In addition, the tortoise coordinate is obtained up to the first order.
\begin{align}
\frac{d\rho_\text{f}}{dx}&=P_\sigma(\rho)-P_\sigma(\rho) B_I(\rho)\epsilon+\mathcal{O}(\epsilon^2),\quad B_I(\rho)=\frac{3D-5}{4}\sigma(\rho-\sigma).
\end{align}
The normalized potential was also determined to be
\begin{align}\label{eq:peaklocationuptofirst01}
\tilde{V}_\text{f}(\rho)&=P_\sigma(\rho)+P_\sigma(\rho)C_I(\rho)\epsilon+\mathcal{O}(\epsilon^2),\quad C_I(\rho)=\frac{3D-13}{12}+\frac{3D-5}{2}\sigma^2-\frac{3(D+1)}{4}\rho\sigma.
\end{align}
The effect from the finite value of $\epsilon$ modifies the location and peak of the effective potential, thus changing the real and imaginary parts of the QNM frequency. According to Eq.\,(\ref{eq:peaklocationuptofirst01}), the peak location is modified to
\begin{align}
\rho_\text{fp}=\rho_\text{p}+\rho_{I\text{p}} \epsilon,
\end{align}
where
\begin{align}
\rho_{I\text{p}}=-\frac{3(D+1)\sigma P_{\sigma\text p}}{8\sqrt{1+3\sigma^2}}=-\frac{(D+1)\left[-1+\sqrt{1+3\sigma^2}+3\sigma^2\left(3+\sqrt{1+3\sigma^2}\right)\right]}{36\sigma\sqrt{1+3\sigma^2}}.
\end{align}
Note that $\rho_{\text{p}I}$ is negative; therefore, the location of the peak moves closer than before by a finite-$\epsilon$ effect. At the peak, the value of effective potential is obtained as 
\begin{align}
\tilde{V}_\text{f}(\rho_\text{fp})=P_{\sigma\text p}+P_{\sigma\text p}C_{I\text p}\epsilon+\mathcal{O}(\epsilon^2),\quad C_{I\text p}\equiv C_I(\rho_{\text p}).
\end{align}
There is no derivative term on $P_\sigma$ because $P'_\sigma(\rho_\text{p})=0$. $C_{I\text{p}}$ contributes to the finite-$\epsilon$ correction. The peak value of the effective second-derivative potential is
\begin{align}
\frac{d^2\tilde V_{\text{f}}}{dx^2}\big{|}_{\rho_\text{f}=\rho_{\text{fp}}}=-2\sqrt{1+3\sigma^2}P_{\sigma\text{p}}^2-2\sqrt{1+3\sigma^2}P_{\sigma\text{p}}^2C_{II\text{p}}\epsilon+\mathcal{O}(\epsilon^2),
\end{align}
where the finite-$\epsilon$ effect is included in
\begin{align}
C_{II\text{p}}=C_{I\text{p}}-\frac{3D-5}{2}\sigma(\rho_{\text p}-\sigma)-\frac{9(D+1)}{8}\frac{P_{\sigma\text{p}}\sigma^2}
{1+3\sigma^2}.
\end{align}
Finally, we can obtain the QNM frequencies with finite-$\epsilon$ corrections using the WKB method. As the effective potential in Eq.\,(\ref{eq:peaklocationuptofirst01}) is real and independent of the frequency, there are two branches. The $(+)$ branch is 
\begin{align}\label{eq:finiteqnmfrequency10}
\operatorname{Re}\tilde{\omega}_{n,+}=&\frac{\sqrt{P_{\sigma\text p}}}{k_\text{eff}}+\frac{\sqrt{P_{\sigma\text p}}}{2}\left(\frac{C_{Ip}}{(D-3)}\sqrt{\frac{3\ell(\ell+D-3)}{2}}+(D-3)\sqrt{\frac{2(1+3\sigma^2)}{3\ell(\ell+D-3)}}\left(n+\frac{1}{2}\right)^2\right)\epsilon^{\frac{1}{2}}+\mathcal{O}(\epsilon^{\frac{3}{2}}),\nonumber\\
\operatorname{Im}\tilde{\omega}_{n,+}=&-\sqrt{P_{\sigma\text p}}
\left(1+3\sigma^2\right)^{1/4}\left(n+\frac{1}{2}\right)\\
&-\sqrt{P_{\sigma\text p}}\left(1+3\sigma^2\right)^{1/4}\frac{1}{2}
\left(
-C_{I\text{p}}+C_{II\text{p}}
-\frac{2(D-3)^2\sqrt{1+3\sigma^2}}{3\ell(\ell+D-3)}\left(n+\frac{1}{2}\right)^2
\right)\left(n+\frac{1}{2}\right)\epsilon
+\mathcal{O}(\epsilon^2).\nonumber
\end{align}
The $(-)$ branch is 
\begin{align}
\operatorname{Re}\tilde{\omega}_{n,-}=-\operatorname{Re}\tilde{\omega}_{n,+},\quad \operatorname{Im}\tilde{\omega}_{n,-}=\operatorname{Im}\tilde{\omega}_{n,+}.
\end{align}
In the finite-$\epsilon$ regime, the QNM frequencies separate from those in the near-ultracold regime into geometric and finite $\ell$ contributions. The leading-order terms in Eq.\,(\ref{eq:finiteqnmfrequency10}) only contain $k_\text{eff}$, which is expressed by a combination of $\epsilon$ and $\ell$. Therefore, the unification with $\epsilon \ll 1$ and $\ell \gg 1$ is a remarkable feature in the ultracold geometry. Furthermore, this type of unification is released into two parts in the finite-$\epsilon$ regime.

\section{Conclusions}

We investigated the correspondence between QNMs and GBFs in the near-ultracold family in four- and higher-dimensional RN--dS black holes by using a massless scalar field. Beginning with the scalar field equations, we redefined the geometric configuration of the black hole in terms of the parameter set $(\epsilon,\sigma)$ that forms the near-ultracold family. Based on this set, we explored the correspondence and QNMs in the near-ultracold family.

The geometric effect becomes important with regard to the correspondence of the ultracold limit $\epsilon\ll1$ and finite $\sigma$. In the ultracold limit, the field equation includes $k_\text{eff}$, which is a combination of $\epsilon$ and $\ell$, and the ultracold limit makes $k_\text{eff}$ tend to zero, similar to the eikonal limit. The details can be found by referring to the QNM frequency in Eq.\,(\ref{eq:qnmunltracold01}). When $k_\text{eff}$ tends to zero, the geometric and eikonal limits move to the same semiclassical regime. Furthermore, the QNM frequencies with finite $\ell$ values are found to be the exact analytic form in the ultracold limit obtained by the WKB method in the leading order of $k_\text{eff}$. Thus, the geometric limit is effectively equivalent to the semiclassical limit.

Particularly, this semiclassical limit plays a major role in the correspondence and also in the null geodesic relationship between QNMs and GBFs. Our QNMs maintain the arbitrary angular quantum number $\ell$ even if the semiclassical limit is considered. Therefore, the corresponding GBFs are also determined using a finite $\ell$. This is a distinguishable feature in the near-ultracold geometry case in higher dimensions. Without the use of the eikonal limit, the correspondence of QNMs and GBFs becomes more precise in the leading order. Furthermore, these make QNMs and GBFs precise to null geodesics. We can relate null circular orbits with QNMs and GBFs in finite $\ell$ cases. Accordingly, the null orbits become more predictable in the near-ultracold geometry case than in other geometries for higher-dimensional RN--dS black holes. Note that we also found the relationship between the derivatives of GBFs and the Lyapunov exponent.

We then moved to two different ways from the QNMs in the ultracold limit. For the $\epsilon\ll 1$ and $\sigma\ll 1$ cases, the geometry becomes the Nariai spacetime; correspondingly, we found the QNM frequency using a general solution to the PT potential without the WKB method. The QNM frequency from the PT potential method is exactly consistent with that of the ultracold limit. Furthermore, we found that the frequency was even for the $\sigma$ reflection. Because of this evenness, there is no linear-$\sigma$ term in Eq.\,(\ref{eq:qnmunltracold01}). We also considered the finite-$\epsilon$ effect. As the finite $\epsilon$ changes the tortoise coordinate, effective potential, and its peak, all of these should be included in the WKB method. The finite-$\epsilon$ QNM frequency is separately contributed by $\epsilon$ and $\ell$. Therefore, the contribution of $\epsilon$ and $\ell$ effects is only unified at the ultracold limit and released in the finite-$\epsilon$ regime.

\vspace{10pt} 

\noindent{\bf Acknowledgments}

\noindent This research was supported by Basic Science Research Program through the National Research Foundation of Korea (NRF) funded by the Ministry of Education (NRF-2022R1I1A2063176) and the Dongguk University Research Fund of 2026.\\

\bibliography{References}
\bibliographystyle{Refsty}
\end{document}